# Optimized Arithmetic Coding for Efficient Data Compression in the Resource-Constrained Internet of Things(IoT)

Vatsala Upadhyay[1], Dr. J. Kokila[2], Dr. Abhishek Vaish[3*]

[1]Department of Information Technology, Indian Institute of Information Technology Allahabad, Prayagraj, 211015, Uttar Pradesh, India. [2]Department of Computer Science and Engineering, Indian Institute of Information Technology Trichy, Tiruchirappalli, 620012, Tamil Nadu, India.
[3]Department of Information Technology, Indian Institute of Information Technology Allahabad, Prayagraj, 211015, Uttar Pradesh, India.

*Corresponding author(s). E-mail(s): abhishek@iiita.ac.in;
Contributing authors: vatsau01@gmail.com; kokilaj@iiitt.ac.in;

**Abstract**

The Internet of Things (IoT) generates vast amounts of heterogeneous data, ranging from sensor readings to log alerts and images, that pose challenges to storage and data transmission in resource-constrained environments. In this context, lossless data compression techniques, like Arithmetic Coding, offer an effective solution owing to their high compression ratio. However, the standard Arithmetic Coding technique is computationally intensive, leading to high memory and processing overhead. This paper proposes an optimized version of Arithmetic coding for the IoT environment that incorporates three improvements using Iterative and Iteration Optimizations for minimizing redundant computations and achieving faster convergence; Principal Component Analysis(PCA) for dimensionality reduction and identifying key features; and lastly, Cardinality reduction for grouping similar probabilities to improve the compression efficiency. The proposed method was evaluated on a dataset of images and demonstrated significant reductions in the time to compress, CPU utilization, and memory consumption, and preserves data integrity as seen through the low RMSE values. The optimized version of the Arithmetic Coding algorithm achieves an impressive compression ratio of 814:1 and 101 ms to compress a single image. This makes the optimized algorithm suitable for real-time applications and resource-constrained environments for efficient data transmission and storage.

## 1 Introduction

The exponential growth in the interconnected digital infrastructure has led to the exponential growth of IoT devices. IoT devices have become an integral component in digital infrastructure due to digital transformation, e.g., Industrial IoT, Medical IoT, SCADA-based systems, and other legacy systems, considering the potential of digital transformation. IoT devices are resource-constrained regarding bandwidth, power, memory, battery, and computational capacity due to their portability features and small sizes[1]. The vast amount of data generated brings about the challenges of data management and storage, security, and efficient transmission[2],[3]. Several techniques have been developed to address these issues, such as blockchain-based solutions that ensure tamper-proof and decentralized data storage, enhancing integrity but requiring significant computational resources[6]. AES (Advanced Encryption Standard) and RSA (Rivest-Shamir-Adleman) algorithms protect sensitive information during trans mission but impose computational overhead on resource-constrained IoT devices[4]. Edge computing minimizes latency and reduces bandwidth usage by processing data locally, although the capabilities of edge devices limit it[5]. Compression algorithms, such as arithmetic coding, optimize data transmission and storage by reducing data size at the cost of decompression overhead. Data compression is pivotal in optimizing data management and enhancing overall system performance in resource-constrained environments like IoT networks. Data compression can mitigate these limitations by minimizing the amount of data stored and transmitted, facilitating more efficient resource utilization, and lowering latency factors critical for real-time communication and remote operations. This reduction helps conserve bandwidth and storage, decreasing energy consumption, which is crucial for IoT devices[1]. Compression techniques minimize data footprints, enable devices with limited memory to store and process larger datasets, extend data retention periods, and enhance system longevity. It also improves energy efficiency, as fewer bits are transmitted and processed, conserving power in battery-operated devices[7]. Data compression plays an important role, especially in the industrial and healthcare sectors, by addressing challenges related to data volume, transmission speed, storage, and energy consumption[7]. The examples given below illustrate the advantages of using data compression in real-time scenarios:

- Real-time Monitoring and Predictive Maintenance: Industrial IoT devices continuously monitor machinery and equipment, generating large datasets from sensors (vibration, temperature, pressure). Compression reduces the generated data size, allowing faster transmission to the cloud platforms for predictive analytics. This minimizes downtime by predicting failures before they occur[7].
- Medical Imaging and Wearable Devices: Medical devices, such as MRI, CT, ultrasound machines, and wearable devices, generate high-resolution images and continuous data that occupy large storage space. Lossless compression ensures that image quality is preserved for accurate diagnosis and enables efficient storage and transmission to healthcare providers for real-time monitoring[1],[7].

The data compression techniques can be broadly categorized into lossless and lossy methods depending on the type of output required (loss or no loss in resultant data):

- Lossless compression: This method ensures that the original data can be perfectly reconstructed from the compressed data. Techniques such as Huffman coding, Run Length Encoding (RLE), Arithmetic Coding, and Lempel-Ziv-Welch (LZW) are commonly used. Lossless compression techniques are helpful in applications where data integrity is of utmost priority, such as healthcare (medical imaging), firmware updates, sensor data, industrial monitoring, etc[1],[9].
- Lossy compression: This method achieves higher compression ratios than lossless compression methods by allowing some loss of data (that is, irreversible, implying that data once lost cannot be retrieved), which may be acceptable in scenarios where exact data precision is not critical. Discrete cosine transform (DCT), wavelet compression, and JPEG (Joint Photographic Experts Group) compression are often used in multimedia data such as images and audio[1].

Several compression algorithms have been adapted and optimized for IoT environments, like the Huffman Coding, a lossless variable-length compression coding technique that assigns shorter codes to more frequent data. This method is effective in scenarios where specific data values appear more frequently, such as sensor data generated from IoT devices. This algorithm is less computationally intensive, making it suitable for IoT devices due to the low power consumption associated with this algorithm[8],[9]. Similarly, Run-length encoding (RLE) is a lossless compression technique that reduces the physical size of repeating characters or patterns, such as sensor readings, in a stable environment. Lempel-Ziv-Welch (LZW) is another lossless compression technique that assigns dictionary references to repeated data occurrences in the input[9]. Similarly, Predictive Coding relies on the previous data values to predict the next value; only the difference between the previous and the next is stored. This technique is efficient for time-series data common in IoT applications in determining the next steps and making informed decisions based on the generated data[10]. Unlike traditional methods like Huffman coding, which encodes each symbol separately, Arithmetic coding is advantageous, especially for data with skewed probability distributions. The algorithm provides higher compression rates than lossless compression algorithms by encoding data into a smaller representation, reducing transmission bandwidth and storage requirements, and is thus effective for textual and multimedia data commonly found in IoT[11].

### 1.1 Arithmetic Coding Compression

Arithmetic coding involves several key steps, wherein the probability for each symbol in the input data is calculated based on its frequency of occurrence in the output or pre-defined probabilities. Then, the interval is subdivided into sub-ranges and refined each time for each symbol in the input. After all the symbols are processed, the final interval is an encoded value - a single floating-point number between 0 and 1 representing the entire sequence of symbols.

Arithmetic coding offers several benefits that mark its suitability for IoT devices and is often considered for data compression due to its efficiency and adaptability in various scenarios, like High Compression Ratio: Arithmetic coding provides higher compression ratios than traditional methods like Huffman coding, thus enabling efficient data storage and fast data transmission required in an IoT environment. Adaptability: Arithmetic coding works well in environments where the probability distribution of data symbols changes dynamically, for example, in IoT data streams[12]. Time and Space Complexity: Huffman Coding involves less complex computations than arithmetic coding, but incurs more time to compute, the order O(n log n)(due to constructing a binary tree). Also, the space required in the case of Huffman Coding is greater than that of arithmetic coding. The time and space complexity incurred for Arithmetic Coding are O(n) and O(1), respectively, where n is the number of symbols. The algorithm's complexity is less than that of Huffman Coding, proving to be a better algorithm.

### 1.2 The Limitations of the Arithmetic Coding Compression Algorithm

While it is seen that there are several advantages of implementing arithmetic coding for compression purposes, it also comes with certain limitations that are addressed below:

1. Computational Complexity and Speed: Arithmetic coding involves complex operations such as high-precision arithmetic and bit-level manipulations, which can be computationally intensive and thus unsuitable in a resource and storage-constrained environment like IoT, according to [13]. This results in slower encoding and decoding speeds than simpler techniques like Huffman coding.
2. Numerical Precision and Overflow: Arithmetic coding requires precise operations on numbers that can grow very large or very small. Maintaining sufficient precision can be challenging and may lead to numerical overflow or underflow in fixed-precision arithmetic environments, as discussed in [14].
3. Error Propagation: A single-bit error can corrupt the entire decoded message, making error correction and detection more challenging than other methods like Huffman coding, where errors are more localized. It was noted in [15] that the global nature of arithmetic coding means errors can have widespread effects on the decoded output, thus requiring robust error correction mechanisms. Any error occurring in an environment like IoT that requires real-time processing and analysis will result in irreversible fatal errors, leading to incorrect results and decisions based on the data obtained.
4. Context Modeling Complexity: The efficiency of arithmetic coding relies heav ily on the effectiveness of the context modeling used to predict symbol probabilities, as discussed in [16], while improving compression rates, adding to the complexity and resource demands of the algorithm.

### 1.3 Our contributions

Based on the analysis of Arithmetic Coding regarding the advantages and disadvantages of its usage in the IoT environment, we propose a modified version of the compression algorithm that aims to optimize the current arithmetic coding algorithm and make it feasible to be deployed for data management in the IoT using the below scientific techniques and achieving the following objectives:

- Reducing the number of iterations incurred to achieve compression by scientific reduction using the techniques of Iterative and Iteration Optimization, Dimensionality reduction using PCA, and Cardinality reduction.
- Clustering similar symbols together to reach convergence faster, further reducing the algorithm's time, space, and computational complexity for efficient processing. • Reducing the dimensionality of data to remove noise and incur a low computational cost, thus further optimizing the compression algorithm.
- Achieving a high compression ratio and faster compression time without compromising the data integrity.
- Improving the existing compression algorithm for applicability across image data types.

The rest of the paper is organized as follows: Section 2 discusses the Literature review done in the area of data compression in the IoT environment and Arith metic Coding and the methodology and shortcomings in the existing works; Section 3 discusses the Experimental setup, dataset, and hardware specifications; Section 4 describes the Results and Analysis obtained from the experiment performed and a comparison with the existing works. Finally, Section 5 discusses the summary of our work and the scope of further improvement in the resultant work that the research fraternity can explore.

## 2 Literature Review

As discussed in the previous section, data compression optimizes storage and transmission in resource-constrained environments like the IoT. Arithmetic coding is a popular and powerful compression technique as it achieves compression by encoding data in fractional intervals[35]. However, the Arithmetic Coding algorithm suffers from certain drawbacks (computationally intensive), making its adoption less feasible for the IoT environment, thus requiring improvement in the existing algorithm. We review the different compression techniques used in the IoT context, focusing on the variations of Arithmetic Coding used in resource-constrained environments and highlighting the pros and cons of the techniques used.

A lightweight algorithm to reduce the number of data transmissions in IoT sensor networks has been proposed by the authors in [17], achieving significant data size reduction while maintaining data integrity. This paper proposes a Compression-Based Data Reduction (CBDR) technique to optimize energy consumption in IoT sensor networks. The method combines lossy Symbolic Aggregate Approximation (SAX) quantization with lossless Lempel-Ziv-Welch (LZW) compression. SAX reduces the dynamic range of sensor data readings, and the LZW compression further compresses the quantized data. It reduces the amount of transmitted data and processing overhead. The effectiveness of CBDR in conserving energy and achieving high compression ratios compared to traditional methods demonstrates scalability using real-world IoT data. Additionally, the 2-stage compression process incurs high computational overhead, thus making it less suitable for usage in a resource-constrained environment like IoT

The authors in [18] introduce the Tiny Anomaly Compressor, a novel data compression solution for IoT environments. The approach efficiently compresses data by focusing on anomaly detection. The algorithm dynamically uses data eccentricity to identify and compress less critical data, reducing the data footprint without significantly affecting downstream tasks. The approach targets TinyML-based systems to enhance energy efficiency and storage utilization in IoT devices, enabling real-time processing. It enhances real-time data processing for TinyML applications and dynamically adapts to varying IoT data patterns. It also minimizes the energy consumption and storage demands of resource-constrained IoT devices. The method requires careful tuning of eccentricity parameters. This may lead to information loss in highly dynamic datasets.

The authors in [19] review various lossless data compression algorithms suitable for IoT applications, focusing on their implementation in edge computing environments. This paper explores lossless compression techniques tailored for mission-critical applications in IoT. The authors emphasize preserving data fidelity in edge computing scenarios with critical latency and accuracy. The method ensures data integrity for mission-critical applications and reduces transmission latency and storage requirements, thus making it suitable for high-stakes environments like healthcare and industrial automation. It is limited to lossless techniques, potentially sacrificing compression ratios. It also incurs higher computational demands on edge devices, thus limiting its scope for wide deployment in the IoT environment.

The proposed technique in [20] addresses the inefficiencies of traditional algorithms on multi-sensor data and demonstrates improvements in real device scenarios. The authors propose a novel IoT data compression technique to enhance energy efficiency in edge machine learning applications. The approach uses adaptive compression algorithms that minimize data redundancy before transmission to edge nodes, effectively reducing energy consumption. The study highlights the use of lossy and lossless compression techniques to balance data fidelity and energy efficiency. Significant energy savings for IoT edge devices and adaptive compression ensure flexibility for various applications. It also supports real-time processing in edge machine learning systems. The method requires fine-tuning for specific data characteristics. Also, the Lossy compression technique may compromise data accuracy in certain cases.

The work in [21] explores lossy compression techniques for IoT sensor data, utilizing spatio-temporal correlations to achieve higher compression ratios. The proposed technique uses predictive models to identify and remove redundant information while retaining critical patterns in the data. This method is especially effective for IoT applications with spatially or temporally related data, such as environmental monitoring. The method exploits inherent correlations in IoT data to achieve high compression rates and is well-suited for IoT applications with predictable data patterns. It may introduce errors in datasets with a weak correlation that can lead to incorrect results critical for real-time processing. Also, the method incurs computational overhead for identifying and processing spatiotemporal dependencies.

The authors in [22] introduce a novel adaptive compression scheme designed for smart metering applications in IoT, efficiently handling multivariate data streams. The method combines dimensionality reduction and multivariate correlation analysis to minimize redundant data while maintaining essential information. The approach is tailored for efficient transmission in bandwidth-constrained environments. The method maintains essential data for analytics while reducing storage and transmission needs and is specifically designed for smart metering systems. The method is limited to smart metering use cases, thus restricting its applicability across varied data types. Also, the method incurs a high initial computational cost for multivariate analysis.

The authors in [23] proposed a new compression algorithm, RAKE, for low-speed microcontrollers in IoT devices. RAKE is a lightweight, lossless compression algorithm that focuses on simplicity and efficiency, making it suitable for low-power devices with limited computational resources. RAKE uses a rule-based approach to identify and compress recurring data patterns. This lightweight and energy-efficient algorithm is ideal for resource-constrained IoT devices. It ensures lossless compression by preserving data integrity. The algorithm is less effective for complex or irregular datasets and lacks adaptability for dynamic IoT environments.

In [24], the authors introduce a compression technique tailored for smart farming to reduce data transmission complexity without compromising data integrity. Reduces data transmission and storage costs. It may not generalize well to other IoT domains. The framework uses adaptive compression techniques to reduce IoT device data trans mission costs and energy consumption. It focuses on integrating edge and cloud computing to balance efficiency and scalability, and integrates well with cloud-based architectures for scalability. The algorithm faces latency issues for real-time applications and incurs computational overhead at the cloud layer, making it less suitable for adoption in resource-constrained environments.

The authors in [25] propose a 2-tier data reduction approach, combining simple compression methods at the sensor node with complex techniques at the aggregation level to minimize data transmission in IoT sensor networks. The $1^{st}$ tier involves local pre-processing and compression at the sensor level, while the $2^{nd}$ tier integrates data aggregation at the gateway. This hierarchical approach reduces transmission overhead and extends the lifetime of IoT networks. The proposed method helps reduce energy consumption and prolongs sensor battery life by combining local and centralized data processing for efficient data reduction, but it increases complexity in deployment and may face challenges with real-time data requirements critical for the IoT.

In [26], the authors investigate lossy compression algorithms for sensor data, optimizing compression based on the signal characteristics in IoT environments. By analyzing different signal types and patterns, the authors identify optimal compression strategies tailored to specific data characteristics, improving compression efficiency without losing critical information. The method has limited generalizability for complex, multi-sensor datasets and requires detailed signal analysis, increasing preprocessing time not beneficial in real-time environments.

The authors in [27] introduce a lossless compression scheme to reduce EEG data size before transmission to cloud platforms in IoT environments. The method ensures high data fidelity, critical for medical applications, while minimizing transmission costs by leveraging fog computing for local processing and compression. It also guarantees data integrity with lossless compression, which is vital for healthcare but is limited to EEG data and may require adaptations for other data types. Also, it relies on fog infrastructure, which may not be universally available.

A sensor-agnostic data compression technique has been proposed in [28] that can be applied universally across various IoT platforms. The technique uses a hybrid model combining statistical and machine learning methods to compress data while maintaining cross-sensor compatibility. The method finds its usefulness in its universal applicability across diverse IoT sensors. It also balances compression performance with generalizability. The hybrid approach increases computational complexity, thereby increasing the computational load on the device. It may not achieve optimal compression for specialized sensors.

The authors in [29] explore integrating data compression techniques with deep learn ing for robust time series classification in IoT applications. The proposed compression technique reduces data volume, and the preserved critical features are used to train deep-learning models for classification tasks. The method also demonstrates robust classification performance for IoT time-series data. It requires high computational power to train deep models. Also, the compression may overlook subtle temporal features in time series, leading to incorrect results and findings.

A novel data compression technique to enhance energy efficiency in IoT solutions has been proposed in [30] by employing a new delta-compression coding technique. The authors aim to reduce energy consumption in IoT devices without compromising data quality by implementing lightweight compression algorithms and emphasizing energy efficiency for resource-constrained devices. Compared to advanced methods, the methods find limited scalability for large, complex datasets and may not achieve high compression ratios.

The review paper in [31] compares various data compression algorithms applied in wireless sensor networks, providing insights into their utility in IoT environments.

In [32], the authors aim to balance data quality and power consumption through their hybrid compression technique, making it suitable for IoT sensor networks. The approach combines lossy and lossless compression methods based on sensor data characteristics to optimize energy efficiency. The compression method reduces power consumption in resource-constrained sensors. The method requires sensor-specific tuning to achieve optimal results and may not generalize well to non-sensor data, i.e., the method is restricted to sensor data type.

A series of enhancements to adaptive arithmetic coding to improve the compression ratio has been proposed for video data in [33]. The work introduced three novel techniques: the range-adjusting scheme, the increasingly adjusting step, and the mutual-learning scheme. The range-adjusting scheme dynamically modifies the frequency table near the current symbol and reduces redundancy. The increasingly adjusting step optimized the convergence by emphasizing recently observed symbols, whereas the mutual-learning scheme enables probabilistic adjustments across neighboring symbols. The proposed approach makes the algorithm adaptive across various video formats and improves the compression performance. The computational overhead associated with frequent probability updates in the range-adjustment scheme presents challenges for IoT devices.

In [34], the authors explored the integration of arithmetic coding and transformer-based architectures for anomalous traffic detection in IoT environments. This hybrid approach takes advantage of the high contextual awareness of transformers and the compression efficiency of arithmetic coding, resulting in more precise anomaly detection. The approach is computationally intensive, posing latency issues in real-time detection scenarios. Also, the resource demands of transformers may limit their deployment in low-power IoT devices.
A combination of the N-gram technique with arithmetic coding has been proposed in [35] to improve the compression of textual data. 5 different text files ranging from 8KB to 100KB were used to evaluate the proposed algorithm's performance. The approach utilizes N-gram analysis for N=1,2,3,4,5 and combines it with the standard arithmetic coding approach. It records the compression and decompression speed, the compression ratio, and the time taken to compress and decompress. The method achieves a high compression ratio, approximately 6:1, by extending to multi-symbol patterns, particularly for large textual datasets. The method is beneficial because of its simplicity in text compression. The technique is limited to text data and has not been tested for image, multimedia, or non-textual formats.
The authors have given a technique tailored for block-based lossless image compression [36]. The approach utilizes mixture models to dynamically adjust probability distributions during encoding, i.e., an adaptive arithmetic coding approach, allowing for a response to varying statistical properties across image blocks. By segmenting images into blocks and applying adaptive modeling, the method achieves high compression ratios without sacrificing image quality. The block-based structure ensures efficient lossless compression, making the technique well-suited for medical imaging, satellite imagery, and other applications requiring pixel-perfect reconstruction. The limitation of this approach lies in its mixture model training that introduces pre-processing overhead, and the high computational demands limit the method's deployment in real-time scenarios like IoT. Also, the block-based strategy may introduce latency when applied to streaming data and continuous feeds.
In [37], the authors proposed a method where the fractional value resulting from arithmetic coding is converted into its equivalent binary form. This binary representation is then segmented into groups of eight bits, each corresponding to a character in the UTF-8. The authors applied their modified arithmetic coding technique to various text datasets to assess its efficacy. The results indicated an improvement in compression rates compared to standard arithmetic coding methods because of the transformation of binary groups into UTF-8 characters, effectively reducing the encoded message length. UTF-8 encoding has been used to ensure the compressed data is compatible with standard text processing systems. The additional binary conversion and segmentation introduce significant computational overhead, impacting its performance in the IoT environment.
The authors in [38] focus on the potential to optimize energy consumption and memory usage within WSNs. The study utilized 4 distinct real-world IoT datasets: temperature (Temp), sea-level pressure (Pressure), stride interval (Stride), and heart rate (BPM). By applying arithmetic coding to these datasets, the authors aimed to assess the technique's effectiveness in preserving the limited resources of sensor nodes. The compression ratios varied across datasets, with the heart rate dataset obtaining a compression ratio of 0.159. Conversely, temperature and pressure datasets resulted in higher ratios of 0.428 and 0.255, respectively, indicating that the effectiveness of arithmetic coding depends upon the nature of the data. The proposed approach optimizes energy conservation and memory through reduced data transmission and storage requirements. This is particularly relevant for IoT environments where frequent data collection and limited power resources require efficient data handling. The authors highlight that compressing sensor data before transmission can significantly reduce the overall energy expenditure in WSNs, thereby increasing the lifespan of IoT devices. While arithmetic coding offers substantial compression gains, the encoding and decoding processes introduce additional computational demands, a limitation of this approach for IoT devices with limited processing capabilities. Also, it was observed that the technique may not uniformly benefit all sensor data types, limiting its generalizability in heterogeneous IoT environments.
The application of Iterative Optimization applies repetitive processes to refine any algorithm's encoding or transformation process to improve the solution progressively. It achieves faster convergence as it observes the threshold from which there will be little to no changes to the solution[39].
To address communication bottlenecks in distributed systems, the authors in [40] have proposed a method of compressed iterates. Compressing updates at each iteration effectively reduces communication costs while maintaining convergence under specific compression schemes. While the results are promising, the approach suffers from errors introduced by high compression, affecting convergence rates in specific scenarios.

In [41], the authors explored an iterative optimization approach using compressed gradient updates to improve distributed learning efficiency. Under compression, the authors established convergence guarantees and highlighted their potential for resource-constrained systems. However, the trade-off between compression-induced errors and convergence rates remains a limitation.

Ebadi and Ebrahimi [42] proposed a progressive iterative approximation technique for video data compression. The method incrementally refines curve fitting, improving compression efficiency with each iteration. However, the algorithm's convergence heavily depends on initial settings, which can limit its robustness and increase the processing time, hindering its adoption in the IoT environment.

A semantic compression algorithm based on parallel iterative optimization to compress large-scale data was proposed in [43], wherein the method leverages semantic relationships within data to achieve high compression ratios and ensure scalability for big data applications. Despite its effectiveness, the computational overhead of semantic analysis poses challenges in real-time implementations like IoT. Wang et al. in [44] proposed a dynamic sparse PCA approach for sensor data compression in virtual metrology. The algorithm adjusts sparsity dynamically, optimizing dimensionality reduction over time. The algorithm can adapt to dynamic data environments and reduce sensor storage requirements, making it feasible for usage in IoT environments, but it incurs increased computational overhead for dynamic adjustments.

Similarly, in [45], the authors investigated the application of PCA to data compression in IoT-enabled Cyber-Physical Systems (CPS). The study proposed a PCA-based compression algorithm optimized for high-volume data streams typical of IoT environments. By reducing the dimensionality of sensor data, the algorithm improves data processing efficiency and storage management while maintaining data integrity for critical applications like anomaly detection and system monitoring. The study emphasizes balancing compression ratio and data loss to support real-time IoT applications. The method reduces storage and communication overhead, which is crucial for IoT devices with limited resources, enabling real-time processing and analysis in CPS environments and maintaining the integrity of critical features necessary for system reliability. However, the high computational requirements for PCA challenge its implementation in real-time scenarios. Its effectiveness depends on the variability and structure of the dataset.

Loia et al. [46] combined fuzzy transforms with PCA to compress electric signal waveforms in smart grid applications. The proposed method targets efficient archiving and interpretation of electric signals while ensuring minimal distortion. PCA reduces dimensionality, and the fuzzy transform enhances signal reconstruction, enabling accurate post-compression analysis. The methodology is particularly effective for compressing repetitive signal patterns, such as those found in electrical grids, reducing storage and transmission demands, and preserving the interpretability of waveforms, which is crucial for fault detection and system analysis. Also, the method is adaptable to various waveform patterns, particularly in power systems. A significant disadvantage of the method is that computational complexity increases with the size of the waveform dataset, thereby limiting its usage in resource-constrained environments like IoT.

Based on the literature survey, Tables 1,2,3 highlight the drawbacks, the technique used, the parameters considered for evaluating the algorithm's performance, the data type, the compression algorithm used, the potential improvement techniques for the drawbacks in the respective work, and the justification for the use of the technique. It can be inferred from the literature survey that:

- Arithmetic Coding offers a high compression ratio but at the cost of computational overhead, making it less suitable for real-time applications and scenarios involving large-scale data. Optimizing the algorithm to balance the compression ratio and computational efficiency is still to be explored. [33],[37].
- Although arithmetic coding has been explored for general data compression, its application remains limited in resource-constrained environments like IoT. Efficient implementation tailored to low-power and low-memory devices is under-researched [34],[38].

- Current implementation of arithmetic coding often focuses on specific data types like text or images. Research is needed to extend the applicability of arithmetic coding to more complex and heterogeneous data[34],[36].
- The energy consumption of arithmetic coding algorithms, especially in IoT and edge devices, is still high[35],[38].
- Arithmetic Coding has challenges when applied to real-time, dynamic, or rapidly changing data streams. Research on adaptive techniques that work efficiently in such scenarios is lacking[34],[36].

**Table 1:Summary of works utilizing arithmetic coding for data compression (*NA = Not Applicable).**

| Technique Used | Parameters Evaluated | Drawbacks | Dataset Used | Improvement Technique | Justification |
|---|---|---|---|---|---|
| Range adjusting and mutual learning schemes [33] | Compression ratio, computational efficiency, compression time | Complex to implement on hardware | Real-world video data | *NA | *NA |
| Arithmetic coding in transformer architectures [34] | Compression rate, accuracy, throughput | High latency for real-time systems | IIoT trafficdata | Iterative, Iteration optimization | Faster processing rates [39] |
| Arithmetic N-gram compression [35] | Compression ratio, encoding/decoding speed | Only suitable for text data | Text datasets | *NA | *NA |
| Mixture models for image compression [36] | Compression ratio, image quality, processing time | High memory usage | Image datasets | PCA, Cardinality | Uses key fea tures [47, 49, 51, 52] |
| Modified arithmetic coding for text [37] | Compression rate, encoding/decoding time | Poor scalabiity | Text datasets | *NA | *NA |
| Arithmetic coding for IoT data [38] | Compression ratio, energy efficiency, complexity | High energy cost | IoT sensor data | PCA, Iterative optimization | Improves convergence [39, 47, 49, 50, 52] |

## 3 Experimental Setup and Methodology

The proceeding section describes the mathematical working of the Arithmetic Coding Compression algorithm, the details of the dataset, and the hardware specification, as well as the justification for the use of the proposed techniques with their mathematical functioning

**Table 2:Summary of works utilizing other data compression techniques in the IoT environment.**

| Type | Technique Used | Parameters Evaluated | Drawbacks | Dataset Used | Improvement Technique | Justification |
|---|---|---|---|---|---|---|
| Lossless | Compression based data reduction [17] | Energy efficiency, latency, scalability | Limited scalability for high data rates | IoT sensor networks | NA | NA |
| Lossy | Evolving TinyML compression algorithm [18] | Compression ratio, accuracy, energy efficiency | High com putational complexity for real-time environ ments | IoT sensor data | Iteration, Iterative optimization | Faster convergence [39] |
| Lossless | Edge lossless compression [19] | Latency, energy efficiency, reliability | Inefficient for bursty traffic | Mission critical IoT | NA | NA |
| Lossy | Energy efficient IoT data com pression [20] | Compression ratio, energy con sumption | Accuracy trade-offs in lossy meth ods | Edge ML datasets | PCA, Car dinality reduction | Unique feature extraction [47–49, 52] |
| Lossy | Spatiotemporal correlation based compression [21] | Compression ratio, accuracy | Limited scalability for dynamic datasets | IoT big data | NA | NA |
| Lossy | Adaptive multivariate compression [22] | Compression ratio, accuracy, energy efficiency | Grows com plex with dimension ality | Smart metering data | NA | NA |
| Lossless | RAKE com pression algorithm [23] | Compression ratio, com putational cost | Not suitable for hetero geneous data | IoT sensor networks | NA | NA |
| Lossy | Compression and MDL for traffic manage ment [24] | Compression ratio, scal ability, energy | Limited to smart agri culture | Agriculture datasets | NA | NA |
| | | effi ciency | | | | |
| Lossless | Two-tier Data reduc tion [25] | Compress ion ratio, trans mission rate | Complex for large net works | IoT sensor data | NA | NA |
| Lossless | Signal based compression [26] | Compress ion ratio, signal accuracy | Fails on dynamic signals | Sensor signal data | NA | NA |
| Lossless | EEG fog enabled compression [27] | Compress ion ratio, latency, energy effi ciency | EEG-only limitation | EEG datasets | NA | NA |
| Lossless | Sensor agnostic IoT com pression [28] | Compress ionratio, scala bility | Poor with hetero geneous sources | General IoTdata | NA | NA |
| Lossy | Time-series compression with DL [29] | Compress ionratio, clas sification accuracy | High energy use | IoT time series data | PCA, Car dinality | Lightweight processing [47–49, 52] |
| Lossy | Energy efficient compression [30] | Compress ionratio, energy effi ciency | Fails for bursty data | IoT sensor data | NA | NA |
| Lossy | WSN data compression [31] | Compress ionratio, com putational complexit y | Too com plex for dense net works | WSN data | NA | NA |
| Hybrid | Hybrid scheme for power reduction [32] | Compress ionratio, energy con sumption | High com putational cost | Biomedical sensor data | PCA, Cardinality | Feature efficient processing [47–49, 52] |

**Table 3:Data compression techniques for IoT using PCA, cardinality reduction, iterative, and iteration optimization techniques (NA = Not Applicable).**

| Compression Type | Technique | Parameters Evaluated | Drawbacks | Data Type | Improvement Technique | Justification |
|---|---|---|---|---|---|---|
| Lossy | Distributed fixed point methods with compressed iterates [40] | Convergen ce rate,comm unication overhead, scalabil ity | Depends on network structure and compression level | Distribute d data | NA | NA |
| Lossy | Gradient descentwith compressed iterates [41] | Gradient sparsity·con vergence speed, computatio nal cost | Accuracy loss due to compression | Optimizatio ndata | PCA, Cardi reduction | Processing necessary features [48], [49], [51], [52] |
| Lossy | Progressive iterative approximation [42] | Compression ratio·iterative efficiency | High computa tional complexity | Video data | Iterative, Iteration optimization | Achieving faster conver gence [39] |
| Lossy | Iterative semantic compression [43] | Compression ratio,parallel processing | Limited to specific semantic features | Large-scal edata | Cardinality | Grouping similarities for faster processing [47], [48] |
| Lossy | Dynamic sparse PCA [44] | Dimensiona lity reduction, data sparsity | High overhead for large datasets | Sensor data | PCA, Cardinali redu | Processing necessary features [47], [48], [49], [52] |
| Lossless | PCA-based data collection [45] | Energy efficiency,co mpression ratio | Needs pre processing for complex data | IoT and CPS data | PCA, Cardi reduction | Processing necessary features [47], [49], [51], [52] |
| Lossy | Fuzzy transform based compression [46] | Energy efficiency,sig nal accuracy | Reduced accuracy for dynamic signals | Electric sig nals | NA | NA |

3.1 Analysis of the Arithmetic Coding Compression

Figure 1 shows the flowchart for working the Arithmetic Coding Compression Algorithm. Let S=$s_1$,$s_2$,...,$s_n$ be the set of symbols, each with a probability P($s_i$), i = 1,2,3. . . .n. • The cumulative probability of $s_i$is defined as in (1):

$$C(s_i) = \sum_{j=1}^{i-1} P(s_j) \quad (1)$$

- For a sequence of symbols X=$x_1$,$x_2$,...,$x_k$, arithmetic coding iteratively narrows down an interval [L, U); L is the lower bound, i.e., 0, and U is the upper bound, i.e., 1.

Encoding process

1. Initialization: The interval is defined as [0,1) for narrowing down the probabilities in this range. Also, the lower and upper limits are defined as L=0 and U=1. 2. Iteration: For each symbol $x_i$in the input, the interval is updated as in (2) and (3):

$$L' = L + [(U - L) * C(x_i)] \quad (2)$$
$$U' = L + [(U - L) * [C(x_i) + P(x_i)]] \quad (3)$$

This process is repeated until the last symbol is processed, yielding a number within [L, U) that uniquely represents the sequence X.

3. Final Encoded value: The final encoded value is any number within the interval [L, U) for the sequence.

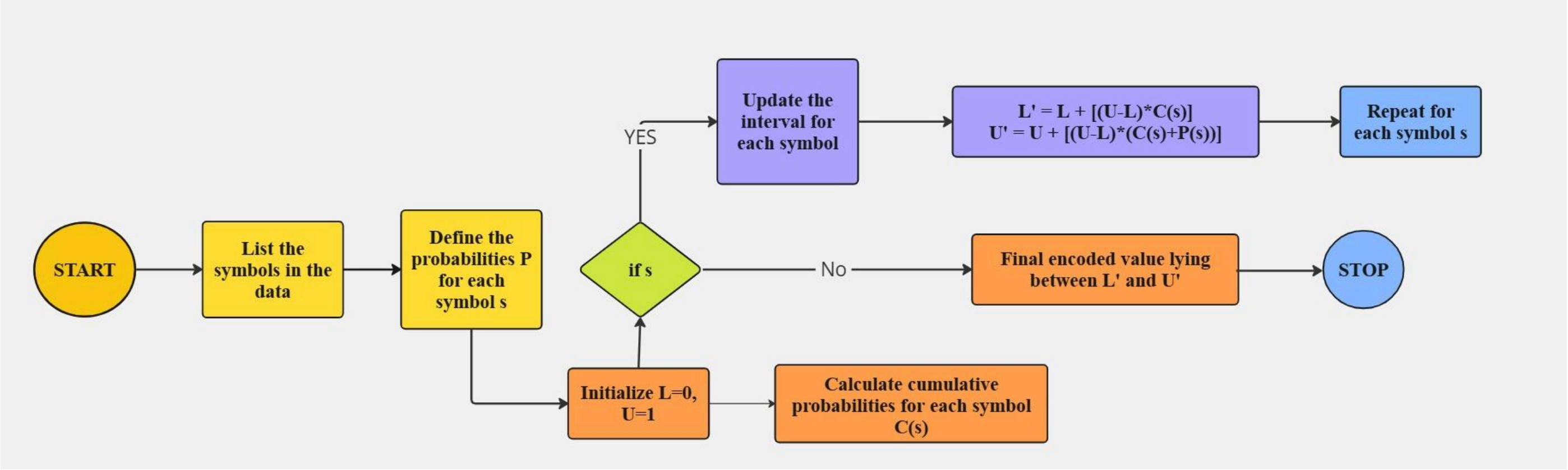

**Fig. 1: Flowchart for the working of the Arithmetic Coding Compression algorithm.**

3.2 The justification for the techniques used, hardware specifications, and dataset details

The proposed method optimizes the existing Arithmetic Coding algorithm and makes it feasible to be deployed in resource-constrained environments like IoT; correspondingly, Figure 2 shows the workflow of our proposed model with the 4 techniques. For our experimental approach, the images were converted to a Grayscale format in a 2-dimensional array; the grayscale conversion has been explained in Algorithm 1 and Figure 3. Converting the image to grayscale improves the compression speed. It reduces the memory consumption and computational load on the device compared to colored images, making it beneficial in resource-constrained environments. Also, the grayscale images contain fewer variations in pixel intensity compared to colored images (in RGB form). Fewer distinct values mean more redundancy, enabling higher compression ratios[53].

1. Iteration Optimization: It improves the iteration process by tuning the parameters so that the computational load can be reduced and memory usage or resources can be optimized, which is beneficial for a resource-constrained environment like IoT. Mathematically, it is defined as stated in (4):

$$x^{k+1} = \varphi(x^k),\ k \geq 0 \quad (4)$$

where $x^k + 1$ = The next solution
$x^k$ = The current solution
k = The no. of iterations

$\varphi$ = The function applied iteratively

For a compression algorithm, (4) can be alternatively represented as in (5):

$$x^{k+1} = f((x^k), D),\ k \geq 0 \quad (5)$$

where D = the original data

f(x) = the function applied iteratively

2. Iterative Optimization converges toward an optimal solution and achieves faster convergence as it observes the threshold from which there will be little to no changes to the solution[39]. A stopping criterion is identified to halt the process, which can help decrease the computational burden on the IoT device in use and resulting in a faster processing time in case of data compression. Mathematically, it can be represented as in (6):

$$min(f(C^k)) \ or \ max(f(C^k)) \quad (6)$$

where f(C) is the function that must be minimized or maximized, depending on the objective, and k is the number of iterations.

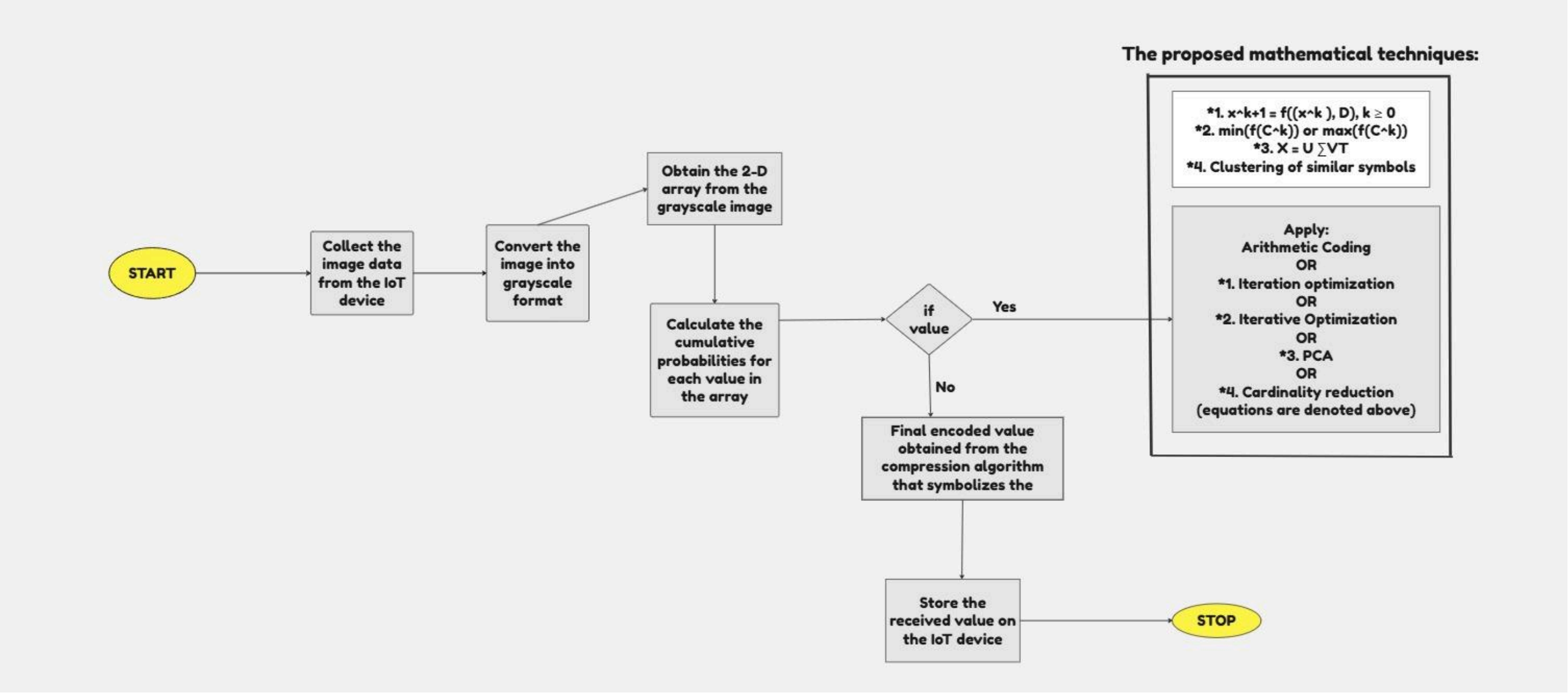

**Fig. 2: Our proposed model and its workflow depicted through a flowchart.**

3. Dimensionality reduction using PCA simplifies the data structure, improving Compression Efficiency by eliminating redundant or less significant features to focus on the most meaningful components. Also, reducing the dimensions minimizes memory and computational resource requirements, thus making it compatible with the constraints of IoT environments. Mathematically, it can be represented as stated in (7):

$$X = U \sum V^T \quad (7)$$

where U and V are the orthogonal matrices, $V^T$is the transpose of V, X is the original high-dimensional matrix, and $\sum$ is the diagonal matrix containing the singular values[49],[50],[51],[52].

4. Cardinality reduction: The method reduces the number of unique symbols in the input data. This helps optimize the probability Distribution, as fewer unique symbols lead to a more efficient calculation of symbol probabilities, reducing computational complexity and achieving a faster convergence rate to improve the algorithm's complexity. Reducing the computational complexity ensures the algorithm performs effectively on devices with limited processing power.

We have created an environment that closely replicates the Raspberry Pi 4 Model B

specifications for testing and programming purposes. The Raspberry Pi 4 Model B is a resource-constrained device with 8GB of RAM, a 64-bit ARM processor, and Thonny IDE (Integrated Development Environment) for Python programming that is pre-installed. Our custom environment mirrors these characteristics and consists of 8GB RAM, a 64-bit processor, and uses the Python programming editor, Jupyter notebook version 7.2.2. Python programming has been chosen as it is considered platform-independent and consists of libraries to support CPU Utilization and memory usage monitoring, graphic visualizations, and the ability to read and process sensor data easily, making it dependable and highly suitable for development in resource-constrained environments. The data type used for the experiments was image data and was divided into 2 categories:

- A single image of data is saved in the desktop system specification of 75.4 KB (612*408 pixels)(.jpg file format).
- A dataset containing 100 images has been created that is compatible with the specifications and constraints of the Raspberry Pi 4. The image sizes vary from 4 KB (kilobytes) to 518 KB. 100 images were downloaded and randomly selected from the Internet with sizes less than 100 MB (.jpg, .jpeg, .png file formats) and saved collectively to meet the specifications of our symbolic resource-constrained Raspberry Pi 4 Model B device.

Table 4 shows the file size distributions for a better understanding of the varieties of file sizes contained in the dataset.

**Table 4: File Size (in KB) distribution in our 100-image dataset.**

| Image | Size | Image | Size | Image | Size | Image | Size | Image | Size |
|---|---|---|---|---|---|---|---|---|---|
| 1 | 510 | 21 | 138 | 41 | 186 | 61 | 101 | 81 | 136 |
| 2 | 139 | 22 | 11 | 42 | 17 | 62 | 129 | 82 | 12.8 |
| 3 | 131 | 23 | 219 | 43 | 7.73 | 63 | 160 | 83 | 47.6 |
| 4 | 11.1 | 24 | 8.64 | 44 | 8.02 | 64 | 15.1 | 84 | 143 |
| 5 | 7.86 | 25 | 104 | 45 | 106 | 65 | 111 | 85 | 10.4 |
| 6 | 18.1 | 26 | 17.2 | 46 | 4.91 | 66 | 195 | 86 | 110 |
| 7 | 358 | 27 | 80.6 | 47 | 214 | 67 | 137 | 87 | 81 |
| 8 | 7.97 | 28 | 18.3 | 48 | 119 | 68 | 233 | 88 | 121 |
| 9 | 11.8 | 29 | 142 | 49 | 59.6 | 69 | 111 | 89 | 12.9 |
| 10 | 46.2 | 30 | 266 | 50 | 6.87 | 70 | 122 | 90 | 111 |
| 11 | 291 | 31 | 4.57 | 51 | 70.7 | 71 | 176 | 91 | 76.2 |
| 12 | 107 | 32 | 118 | 52 | 116 | 72 | 9.64 | 92 | 198 |
| 13 | 40.6 | 33 | 208 | 53 | 157 | 73 | 82.5 | 93 | 109 |
| 14 | 8.98 | 34 | 67.5 | 54 | 153 | 74 | 86.7 | 94 | 170 |
| 15 | 116 | 35 | 140 | 55 | 6.23 | 75 | 97 | 95 | 13.6 |
| 16 | 187 | 36 | 10.2 | 56 | 178 | 76 | 114 | 96 | 172 |
| 17 | 7.63 | 37 | 118 | 57 | 59.8 | 77 | 12.7 | 97 | 130 |
| 18 | 81.9 | 38 | 10.5 | 58 | 148 | 78 | 162 | 98 | 131 |
| 19 | 5.11 | 39 | 11.3 | 59 | 162 | 79 | 108 | 99 | 4.55 |
| 20 | 107 | 40 | 8.54 | 60 | 518 | 80 | 126 | 100 | 4.21 |

### 3.3 The metrics of evaluating a compression algorithm

The metrics considered for evaluating the performance of the compression algorithm are briefly described below and have been divided into Existing and proposed parameters; the existing parameters have been used to compare our results with the existing works:

1. Existing parameters
   (a) Compression ratio: The ratio between the original and compressed sizes for any given data. It is defined mathematically as in (8):

$$Compression\ ratio(C.R.) = Original\ Size/Compressed\ size \quad (8)$$

A higher compression ratio implies that the data size has been significantly reduced, which can be beneficial for scenarios where storage is a constraint. (b) Time to Compress: The total time taken to perform the compression from the start of the algorithm to completion, i.e., how fast an algorithm can generate the desired results. Mathematically, it can be written as stated in (9):

$$timetaken = endtime - starttime \quad (9)$$

Where the starttime and end time are recorded at the start and end of the coding process.

2. Proposed parameters
   (a) Number of Iterations: In arithmetic coding, the number of iterations refers to the number of times the probability interval is refined during the encoding process.
   (b) Memory Consumed: The amount of system memory, i.e., RAM, consumed during the compression process is termed the Memory Consumption, usually measured in MB (Mega Bytes).
   (c) CPU Utilization refers to the amount of resources or work handled by the system's CPU, measured in %. It indicates the percentage of CPU used to perform the task and the amount of load the system's processors handle to run the program.
   (d) Energy Consumption: The amount of energy incurred by the processors, memory, peripheral devices, etc., during the program's execution is referred to as Energy Consumption, measured in W (Watts).
   (e) Probability threshold: A probability threshold indicates the limit after which the compression process doesn't have much effect on the data quality. Determining the probability threshold is useful as it can decrease the system's computational burden and resource consumption.
   (f) Root Mean Square Error (RMSE) quantifies the error between the original and the compressed image in image compression, reflecting the loss of information during the compression process. This metric proves useful for showing the accuracy of the compression algorithm. Mathematically, it is calculated using

the formula given in (10)

$$RMSE = \sqrt{\frac{1}{n}\sum_{i=1}^{n}(x_i - \hat{x}_i)^2} \quad (10)$$

where n = Total no. of pixels in the image
$x_i$ = The intensity of the ith pixel in the original image
$\hat{x}_i$ = The intensity of the ith pixel in the compressed image
A value of RMSE lying between 1 and 10 for image compression indicates little loss of information and the data quality being preserved in the compressed image. For a lossless compression algorithm, the ideal value of RMSE should be 0 [54],[55],[56].

## 4 Results and Analysis

This section discusses the metrics and the values obtained for the experiments performed on image data using Arithmetic Coding and the modified approaches using *Dimensionality reduction, Iterative and Iteration Optimizations, and Cardinality Reduction*. The section has been divided into subparts describing the results obtained on a single image dataset, the metrics obtained on the 100-image dataset, and the comparative analysis of our approach with the existing approaches.

4.1 The experiments performed along with the pseudo-algorithms

The following section briefly describes the experiments performed on the image data, their definitions, and the pseudo-algorithms for each experiment. Algorithm 1 shows the conversion of the image to grayscale, followed by the proposed methods.

1. Standard Arithmetic Coding approach: Shown through Algorithm 2.
2. Arithmetic Coding using Principal Component Analysis (PCA): This method creates eigenvectors from the 2-D array and thus works only on the projected features obtained. Algorithm 3 shows the pseudo-algorithm for the working of PCA in Arithmetic Coding.
3. Arithmetic Coding using Cardinality Reduction applies grouping or merges similar values. This simplifies the data representation, minimizes the range, and reduces the device's computational overhead, as shown through Algorithm 4.
4. Arithmetic Coding using Iterative and Iteration Optimization: This approach modifies the methodology of existing arithmetic coding by applying three methods and the corresponding pseudo-algorithm in Algorithm 5:

a. Sort the probabilities in decreasing order.
b. Grouping similar probabilities in a group to reduce the iterations.

Algorithm 1: Conversion to Grayscale

Input: image_path (path to the image file)
Output: Grayscale image as a numpy array
START;
load_image (image_path);
begin
  Open the image from image_path;
  Convert the image to grayscale;
  Transform the grayscale image to a numpy array;
  return *grayscale image array*;
end
STOP;

Algorithm 2: Standard Arithmetic Coding Compression Algorithm Input:
Sequence of length $N$, Probability table $P$
Output: Encoded value $v$ in $[l, h)$
START;
Initialize $l = 0$, $h = 1$ (low and high bounds);
for $i = 1$ *to* $N$ do
  Symbol $s = S[i]$;
  range $= h - l$;
  $h = l +$ range $\times$ cumulative_probability($s + 1$);
  $l = l +$ range $\times$ cumulative_probability($s$);
Encoded value $v \in [l, h)$;
STOP;

Algorithm 3: Arithmetic Coding with PCA

Input: Data $D$ (matrix of symbols)
Output: Encoded value $v$ in $[l, h)$
START;
Apply PCA to $D$:;
  Compute the covariance matrix of $D$;
  Find eigenvectors and eigenvalues;
  Select the $k$ largest eigenvectors;
  Transform $D$ to a new $k$-dimensional space;
Initialize $l = 0$, $h = 1$;
for *each transformed symbol s in D* do

range = $h - l$;
$h = l +$ range × cumulative probability($s + 1$);
$l = l +$ range × cumulative probability($s$);

Encoded value $v \in [l, h)$;
STOP;

Algorithm 4: Arithmetic Coding with Cardinality reduction
Input: Sequence $S$, Threshold $T$
Output: Encoded value $v$ in $[l, h)$
START;
Initialize $l = 0$, $h = 1$;
for $i = 1$ to *length(S)* do
if $S[i]$ *is similar to* $S[i-1]$ *(within T)* then
Group $S[i]$ with $S[i-1]$;
else
Add $S[i]$ as a new symbol;

Calculate probabilities for grouped symbols;
for *each symbol s in the grouped sequence* do
range = $h - l$;
$h = l +$ range × cumulative probability($s + 1$);
$l = l +$ range × cumulative probability($s$);

Encoded value $v \in [l, h)$;
STOP;

3. Setting a probability threshold to set stopping criteria.

Algorithm 5: Arithmetic Coding Compression using Iterative and Iteration Optimization
Input: Sequence $S$, Threshold $T$
Output: Encoded value $v$ in $[l, h)$
START;
Sort probabilities in decreasing order $S[i]$;
Initialize $l = 0$, $h = 1$;
for $i = 1$ *to length*($S$) do
if $S[i]$ *is similar to* $S[i-1]$ *(within T)* then
Group $S[i]$ with $S[i-1]$;
else
Add $S[i]$ as a new symbol;

Calculate probabilities for grouped symbols;
for *each symbol s in grouped sequence* do
Calculate range = $h - l$;
$h = l +$ range × cumulative probability($s + 1$);
$l = l +$ range × cumulative probability($s$);

Encoded value $v \in [l, h)$;
STOP;

```
Processing image 1/100: D:/resp_images\image001.jpg

--- Grayscale Image Details ---
Image Path: D:/resp_images\image001.jpg
Dimensions: 800x600
Grayscale Array (Pixel Intensities):
[[104 103  89 ...  86 120 109]
 [ 98 105  95 ... 116 107  94]
 [107  99 103 ...  93  83  89]
 ...
 [ 72  64  69 ...  95  88 102]
 [ 67  64  69 ...  88  87  90]
 [ 68  73  66 ...  92  95  88]]
Processing image 2/100: D:/resp_images\image002.jpg
Processing image 3/100: D:/resp_images\image003.jpg
Processing image 4/100: D:/resp_images\image004.jpeg
Processing image 5/100: D:/resp_images\image005.jpeg
Processing image 6/100: D:/resp_images\image006.jpeg
Processing image 7/100: D:/resp_images\image007.jpg
Processing image 8/100: D:/resp_images\image008.jpeg
```

**Fig. 3: An example of the grayscale image dimensions obtained for the colored input image.**

4.2 Results obtained for single image data

Table 5 shows the metrics recorded for a single image data, and the correspond ing graphs in Figures 4,5,6,7 show the comparison between the standard approach to Arithmetic Coding, the modified approach using PCA and cardinality reduction, respectively, through the metrics of Compression ratio, Time to compression, the mem ory consumption, the no. of iterations incurred, the RMSE, and the CPU utilization incurred. It can be inferred from Table 5 and Figures 4,5,6,7 that:

- The compression ratios achieved through the three approaches are high because the image had been reduced to grayscale, and then the compression algorithm was run on the image.
- Further, while applying PCA, 50% of the components were retained, resulting in a higher compression ratio with little loss of information. Retaining too many components didn’t result in a higher compression ratio, while keeping too few components will result in a huge loss of information.
- Also, while applying cardinality reduction, the no. of levels determines the compression scale. The no. of levels determines the value of the compression ratio - reducing by a large no. of levels will result in data precision loss, whereas a reduction by a smaller no. of levels will result in a lower compression ratio. For our experimental approach, the no. of levels has been set to 32 for a 256-bit grayscale image.
- The RMSE value obtained using the standard Arithmetic Coding is very high, indicating a huge loss of information in the compressed data.
- The RMSE value obtained using the modified approach, i.e., using PCA and cardinality, was approximately 10, indicating a minimal loss in the data integrity, thus proving its usefulness in the compression.
- The time taken to compress the data is significantly very low and indicates that the algorithm is efficient in terms of time complexity.

- The amount of RAM required to run the algorithms is very low, indicating that the algorithm utilizes fewer system resources for its operation and implying its efficiency in terms of space complexity and suitability for adoption in resource-constrained environments.
- The CPU Utilization is higher in the PCA approach because of the extra preprocessing done - calculation of eigenvectors and projection, and then running the algorithm resulted in high CPU utilization, which may not be suitable in a resource-constrained environment.
- The no. of iterations incurred is directly dependent on the no. of pixels in the image. For the PCA approach, the no. of iterations will be lower because of the preprocessing and retention of unique features of the image.

**Table 5: Values recorded when Arithmetic Coding compression is performed on a single image.**

| Parameters | Standard Arithmetic Coding | Arithmetic Coding + PCA | Arithmetic Coding + Cardinality reduction |
|---|---|---|---|
| Original Size (bytes) | 77,296 | 77,296 | 77,296 |
| Compressed Size (bytes) | 95 | 102 | 96 |
| Compression Ratio | 813:1 | 758:1 | 805:1 |
| Time to Compress (sec) | 0.1019 | 0.0654 | 0.1164 |
| No. of Iterations | 249,748 | 83,291 | 249,748 |
| Memory Consumed (MB) | 0.0039 | 1.1250 | – |
| CPU Utilization (%) | 44.5 | 148.80 | 56.80 |
| Root Mean Square Error (RMSE) | 85.1688 | 10.2947 | 10.239 |

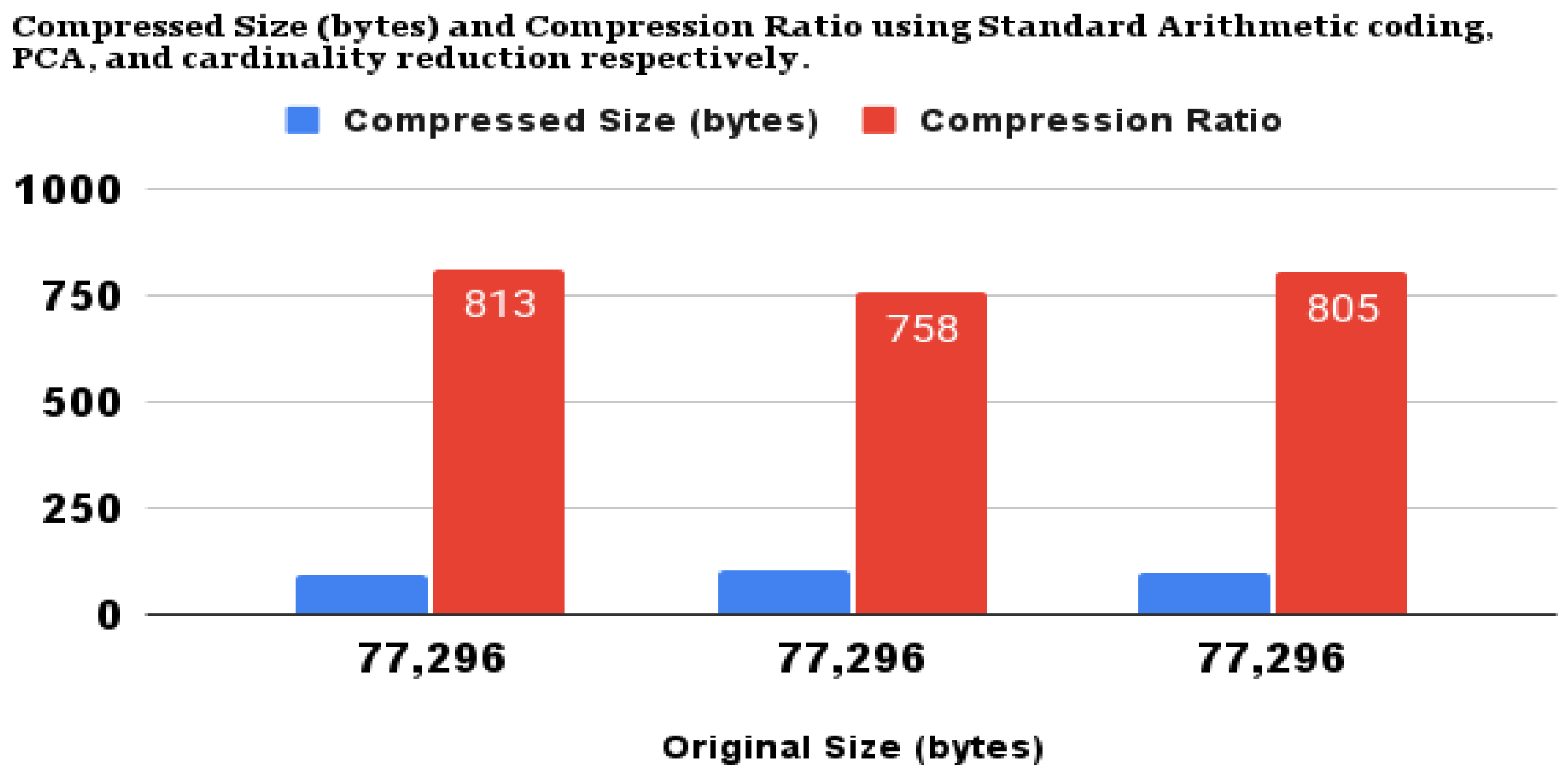

**Fig. 4: The graphic visualization for the compression metrics recorded using Arithmetic Coding (AC), AC+PCA, and AC+cardinality reduction for a single image.**

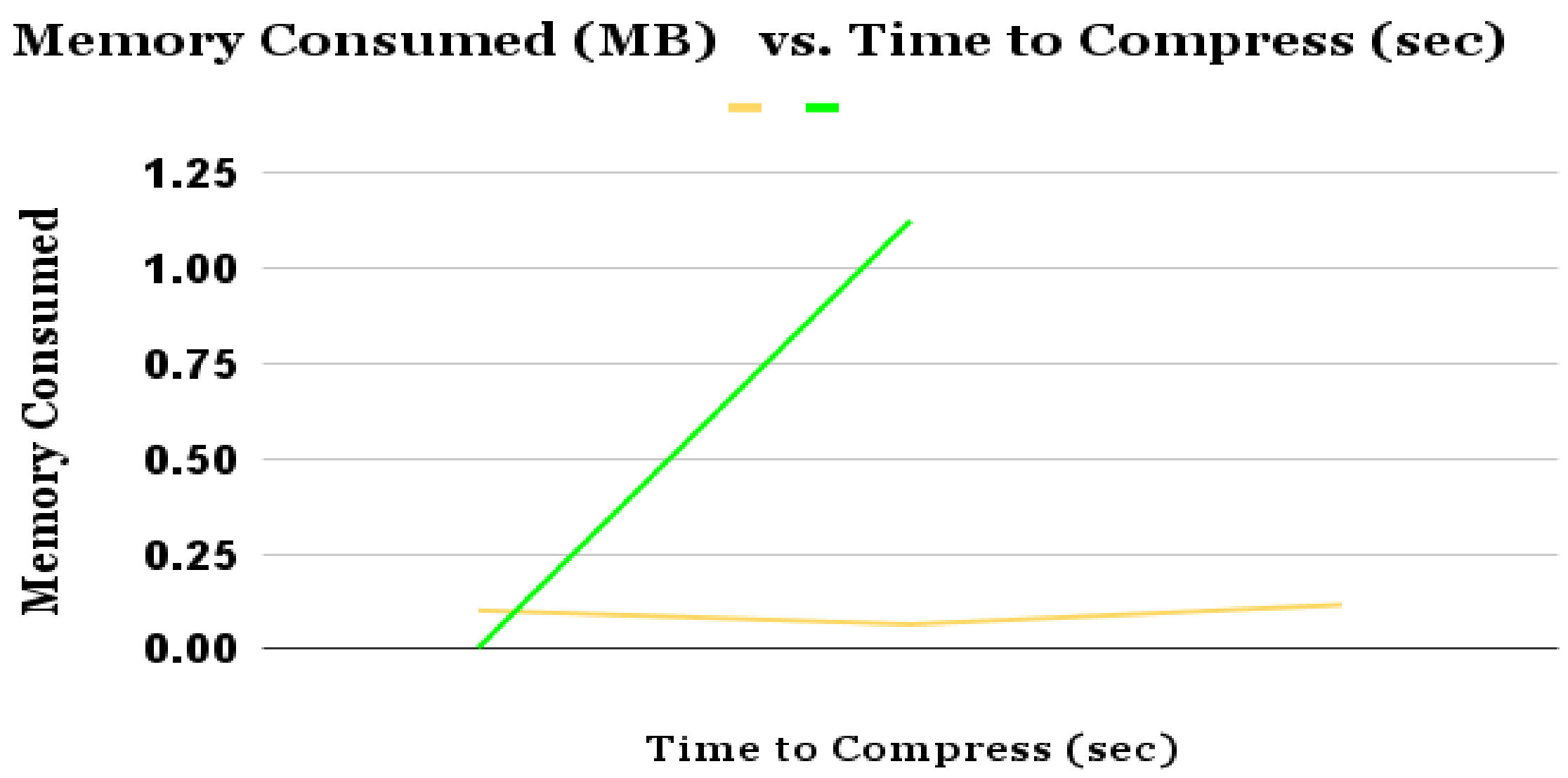

**Fig. 5: The graphic visualization for the memory consumption and time to compress recorded using Arithmetic Coding (AC) and AC+PCA for a single image.**

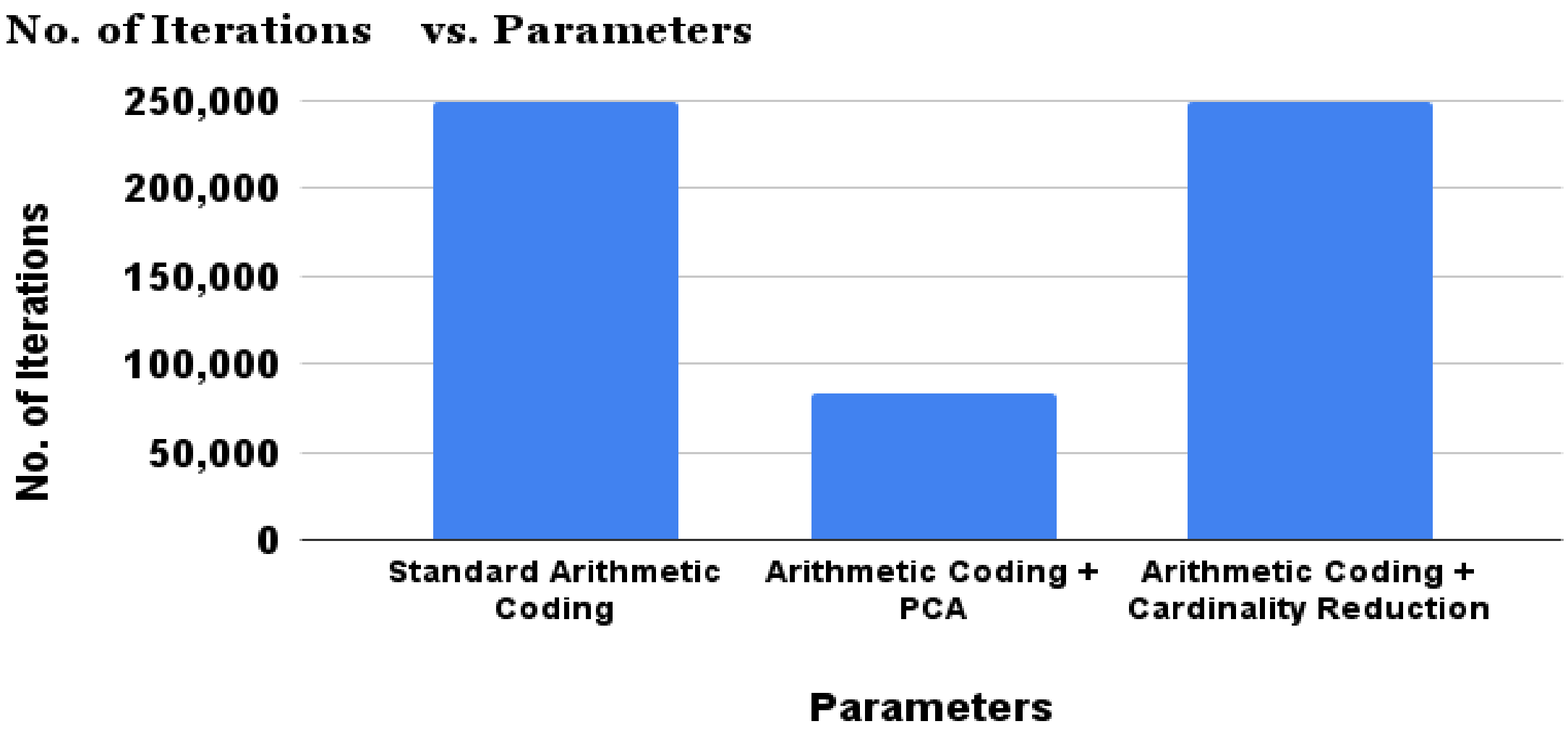

**Fig. 6: The graphic visualization for the number of iterations recorded using Arithmetic Coding (AC), AC+PCA, and AC+cardinality reduction for a single image.**

Table 6 shows the values recorded when Arithmetic Coding is performed on a single image using the Iterative and Iteration optimization approaches derived from Section 4.1 d. Figures 8 and 9 show the results pictorially through bar graphs for the RMSE, CPU utilization, and the compression ratio obtained. It is observed from Table 6 and Figures 8,9 that:

- The compression ratio is high for all 4 threshold values, i.e., n=3,4,5,6, due to converting the image to a grayscale format and then applying compression, thus significantly reducing the data size. The optimization approach fine-tunes the algorithm and results in a high compression ratio.
- The time taken to compress is slightly greater than the standard approach due to the extra preprocessing done (grouping, sorting, thresholding).
- The RMSE obtained is approximately 10, implying a minimal loss in the data integrity in the compressed data, and is acceptable for an image.
- The CPU Utilization is considerably high because the algorithm works on extra processes - sorting, grouping, and restricting the probability thresholds. All these processes incur extra resources at the cost of a high compression ratio, which may not fit in an IoT environment.

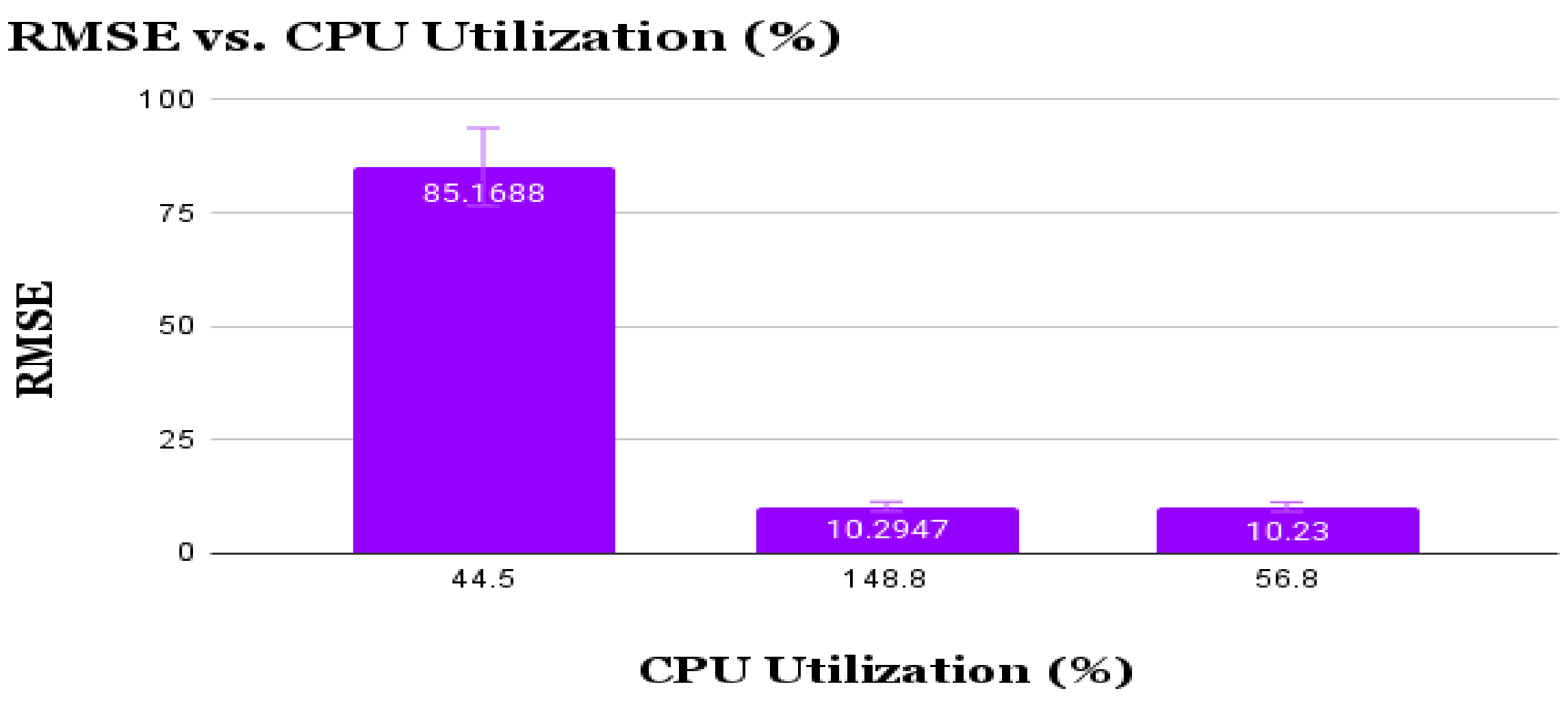

**Fig. 7: The graphic visualization for the CPU utilization and RMSE recorded using Arithmetic Coding (AC), AC+PCA, and AC+cardinality reduction for a single image.**

**Table 6: Metrics recorded for Arithmetic Coding using the Iterative and Iteration Optimization (group size = 6 for similar symbols).**

| Parameters / Probability Thresholds | n=3 | n=4 | n=5 | n=6 |
|---|---|---|---|---|
| Original Size (bytes) | 77,296 | 77,296 | 77,296 | 77,296 |
| Compressed Size (bytes) | 96 | 95 | 96 | 95 |
| Compression Ratio | 805:1 | 814:1 | 805:1 | 814:1 |
| Time to Compress (sec) | 0.1769 | 0.1679 | 0.1943 | 0.1639 |
| Iterations in Compression | 247,949 | 247,949 | 247,949 | 247,949 |
| Root Mean Square Error (RMSE) | 10.2236 | 10.2291 | 10.2301 | 10.2399 |
| CPU Utilization (%) | 99.9 | 99.9 | 100.1 | 99.9 |

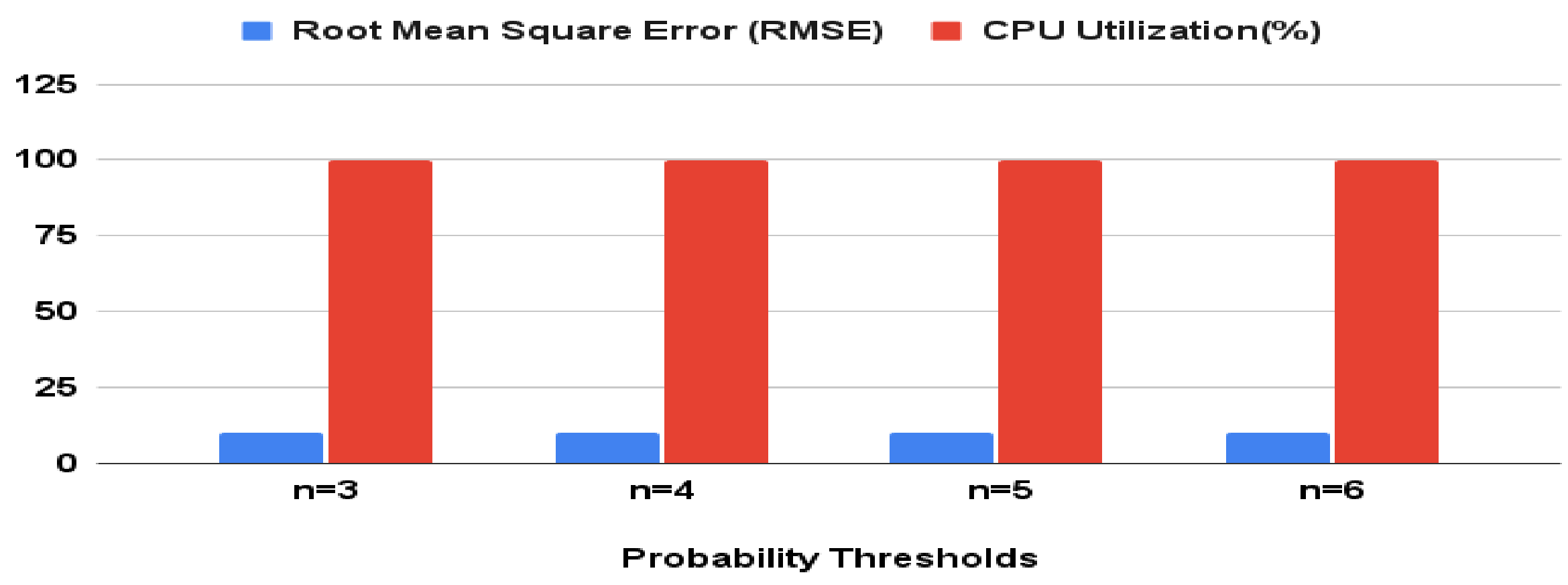

**Fig. 8: Graph showing RMSE and CPU utilization obtained for different probability thresholds for a single image using optimization.**

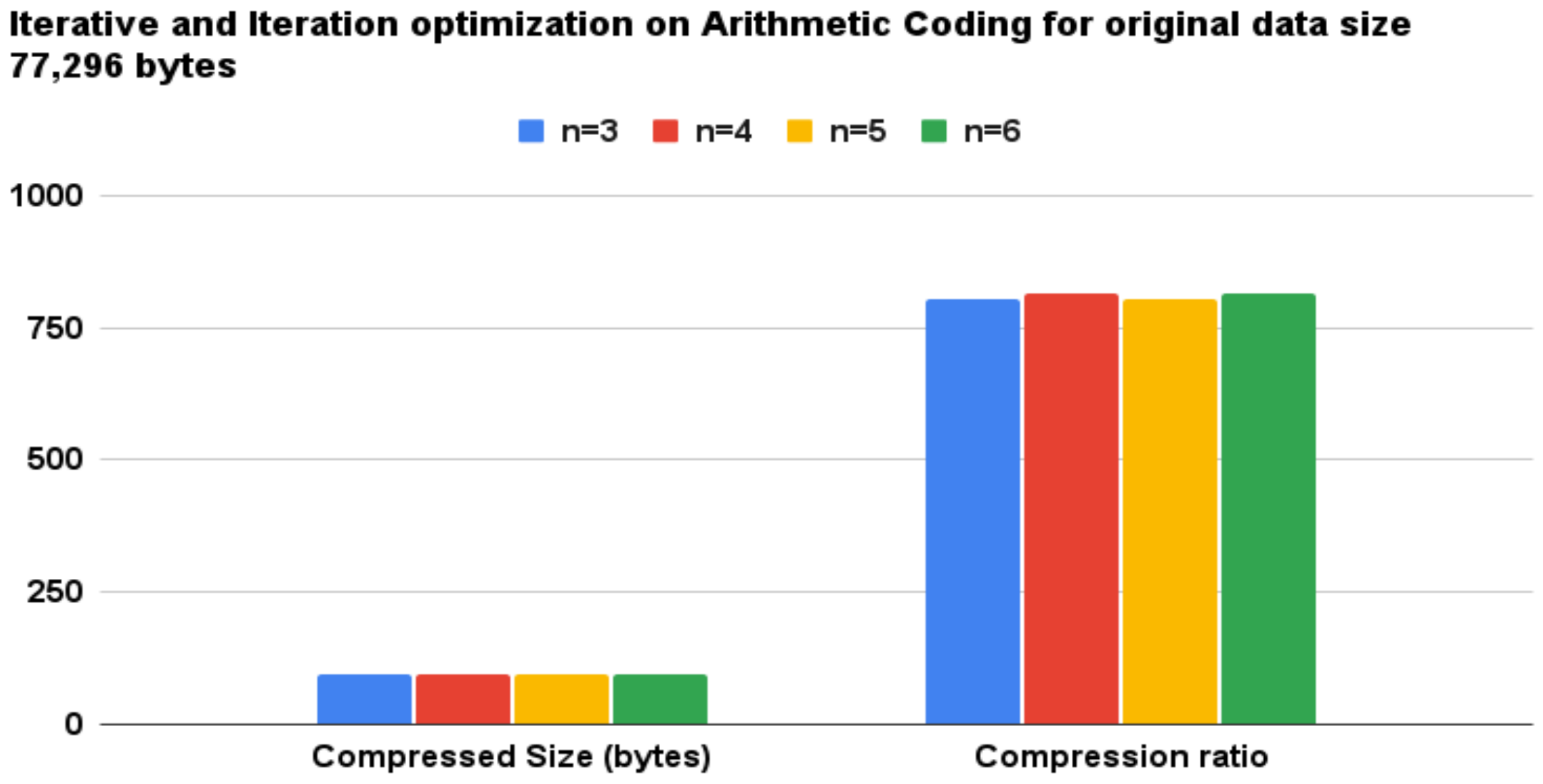

**Fig. 9: Graph showing the compression metrics for different probability thresholds using optimization on a single image.**

4.3 Results obtained for 100-image dataset

Table 7 shows the metrics recorded for the 1st 10 dataset images when Standard Arithmetic Coding is performed. It is inferred from Table 7 that:

- The compression ratio achieved using the standard approach is high and depends on the image features.
- The time taken to compress across all the images is very low, thus proving its efficiency in terms of time complexity.
- The RMSE values are greater than 10, indicating a huge information loss in the compressed data, and are unacceptable for an image.

**Table 7: Arithmetic Coding Metrics for the First 10 Images of the Dataset**

| Image Index | Compression Ratio | RMSE | Time (s) | Original Size(bytes) | Compressed Size (bytes) |
|---|---|---|---|---|---|
| image001.jpg | 2979:1 | 59.07 | 0.0431 | 65536 | 22 |
| image002.jpg | 3120:1 | 90.33 | 0.0262 | 65536 | 21 |
| image003.jpg | 2979:1 | 80.90 | 0.0333 | 65536 | 22 |
| image004.jpeg | 2979:1 | 70.42 | 0.0290 | 65536 | 22 |
| image005.jpeg | 2979:1 | 63.24 | 0.0221 | 65536 | 22 |
| image006.jpeg | 3120:1 | 80.30 | 0.0240 | 65536 | 21 |
| image007.jpg | 3120:1 | 66.17 | 0.0223 | 65536 | 21 |
| image008.jpeg | 2979:1 | 76.90 | 0.0222 | 65536 | 22 |
| image009.jpeg | 2979:1 | 95.33 | 0.0590 | 65536 | 22 |
| image010.jpg | 2979:1 | 103.23 | 0.0212 | 65536 | 22 |

Table 8 and Figures 10 and 11 show the values recorded using Iterative and Iteration Optimization on the dataset's 1st 10 images. There are 2 classes of images - one with 480000 bits and the other with 50246 bits. This is because the images have been converted to grayscale and converted to bits to make the processing faster - the conversion took place using the Pillow library of the Python programming language, which converted it implicitly.

This has been computed as given below:

1. The image dimensions lie in 2 classes: 800*600 pixels and 259*194 pixels. 2. All the images have been tabulated based on their pixel values, thus yielding 480000 pixels and 50246 pixels for computation purposes.

It can be observed from Table 8 and Figures 10, 11 that:

- The compression ratio achieved is high due to the Iteration and Iterative optimization approaches, and is also dependent on the image features.
- The time taken to perform compression was drastically reduced because of the Iteration Optimization algorithm’s convergence and early stopping criteria; the algorithm is optimized for time complexity.
- The memory consumption and CPU utilization are very low and indicate that the modification resulted in early convergence and low requirement of system resources, thus suitable for an IoT.
- Also, it can be observed that the ideal probability threshold is 3, where optimal values are achieved for all the metrics - Time to compress, CPU Utilization, Compression ratio, and the memory consumed during compression. This is another indicator for the stopping criteria for the Arithmetic Coding compression, and an advantage that the system will utilize fewer resources at this stage. Beyond that threshold, there will be a spike in the metrics related to time, CPU, and memory usage that is not beneficial for the IoT environment due to resource constraints.

Table 9 shows Arithmetic Coding and Cardinality reduction on 10 dataset images. It can be observed from Table 9 that:

• A high compression ratio is achieved for all the images, depending on the image features.
• The time taken to perform compression for individual images is very low because of the decrease in the pixel range values through cardinality reduction. • The CPU utilization is low for some images, thus requiring fewer resources and less overhead on the device—the conversion to grayscale results in low CPU utilization. Also, the features in the image directly affect CPU Utilization - an image rich in features will incur more resources than a low-feature image.
• RMSE values are very high, greater than 10, indicating a huge loss in the data integrity in the compressed data.

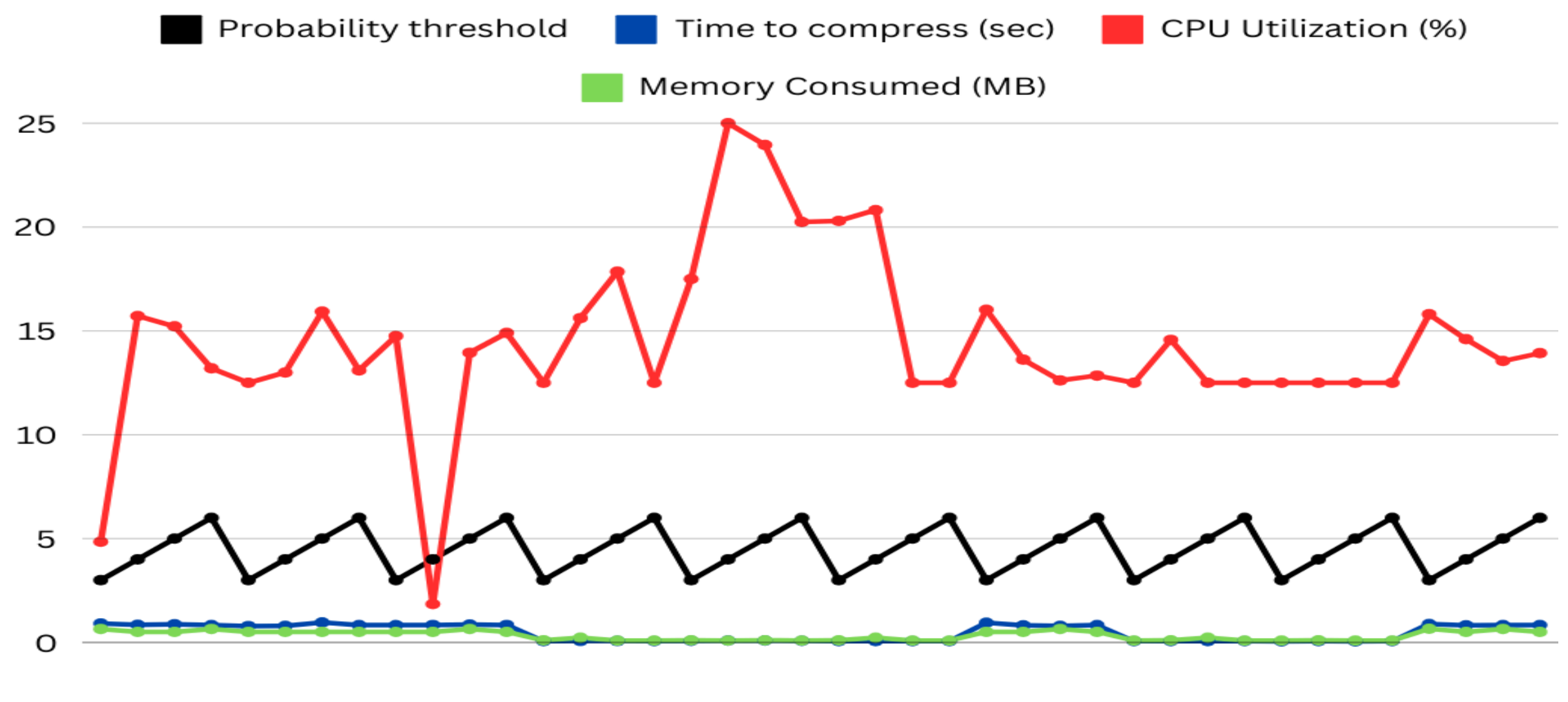

**Fig. 10: The time to compress, the CPU utilization, and memory consumption recorded using the modified arithmetic coding for 10 images.**

4.4 Comparison of the results with the existing approaches

Table 10 compares our approach with the existing approaches on Arithmetic coding in terms of Compression Ratio and the Time to compress for a single image, indicating that the proposed approach outperforms the existing approaches (inferred from Table 5). The Compression ratio achieved is relatively higher because of the use of grayscale conversion on the image. The time taken to compress is reduced due to conversion to grayscale, efficiently decreasing the iteration logic to achieve faster results.

**Table 8: Modified Arithmetic Coding: Metrics for the First 10 Images with Varying Probability Thresholds**

| Image | n | Original Size (px) | Compressed Size (px) | Ratio | Time (s) | CPU (%) | Memory (MB) |
|---|---|---|---|---|---|---|---|
| 1 | 3 | 480000 | 84 | 5714:1 | 0.91 | 4.85 | 0.65 |
| | 4 | 480000 | 84 | 5714:1 | 0.85 | 15.72 | 0.51 |
| | 5 | 480000 | 84 | 5714:1 | 0.87 | 15.22 | 0.51 |
| | 6 | 480000 | 84 | 5714:1 | 0.83 | 13.20 | 0.65 |
| 2 | 3 | 480000 | 85 | 5647:1 | 0.79 | 12.50 | 0.51 |
| | 4 | 480000 | 85 | 5647:1 | 0.80 | 13.00 | 0.51 |
| | 5 | 480000 | 85 | 5647:1 | 0.96 | 15.93 | 0.51 |
| | 6 | 480000 | 85 | 5647:1 | 0.83 | 13.10 | 0.51 |
| 3 | 3 | 480000 | 86 | 5581:1 | 0.83 | 14.75 | 0.51 |
| | 4 | 480000 | 86 | 5581:1 | 0.84 | 1.85 | 0.51 |
| | 5 | 480000 | 86 | 5581:1 | 0.86 | 13.95 | 0.65 |
| | 6 | 480000 | 86 | 5581:1 | 0.84 | 14.90 | 0.51 |
| 4 | 3 | 50246 | 86 | 584:1 | 0.06 | 12.50 | 0.10 |
| | 4 | 50246 | 86 | 584:1 | 0.07 | 15.62 | 0.23 |
| | 5 | 50246 | 86 | 584:1 | 0.07 | 17.85 | 0.09 |
| | 6 | 50246 | 86 | 584:1 | 0.06 | 12.50 | 0.09 |
| 5 | 3 | 50246 | 86 | 584:1 | 0.08 | 17.50 | 0.10 |
| | 4 | 50246 | 86 | 584:1 | 0.09 | 25.00 | 0.09 |
| | 5 | 50246 | 86 | 584:1 | 0.09 | 23.95 | 0.10 |
| | 6 | 50246 | 86 | 584:1 | 0.08 | 20.25 | 0.09 |
| 6 | 3 | 50246 | 86 | 584:1 | 0.06 | 20.30 | 0.10 |
| | 4 | 50246 | 86 | 584:1 | 0.06 | 20.82 | 0.23 |
| | 5 | 50246 | 86 | 584:1 | 0.06 | 12.50 | 0.09 |
| | 6 | 50246 | 86 | 584:1 | 0.06 | 12.50 | 0.09 |
| 7 | 3 | 480000 | 80 | 6000:1 | 0.95 | 16.02 | 0.51 |
| | 4 | 480000 | 80 | 6000:1 | 0.82 | 13.62 | 0.51 |
| | 5 | 480000 | 80 | 6000:1 | 0.80 | 12.62 | 0.65 |
| | 6 | 480000 | 80 | 6000:1 | 0.83 | 12.85 | 0.51 |
| 8 | 3 | 50246 | 86 | 584:1 | 0.06 | 12.50 | 0.09 |
| | 4 | 50246 | 86 | 584:1 | 0.06 | 14.57 | 0.10 |
| | 5 | 50246 | 86 | 584:1 | 0.06 | 12.50 | 0.23 |
| | 6 | 50246 | 86 | 584:1 | 0.06 | 12.50 | 0.09 |
| 9 | 3<br>4<br>5<br>6 | 50246<br>50246<br>50246<br>50246 | 85<br>85<br>85<br>85 | 591:1<br>591:1<br>591:1<br>591:1 | 0.05<br>0.06<br>0.05<br>0.06 | 12.50<br>12.50<br>12.50<br>12.50 | 0.09<br>0.10<br>0.09<br>0.09 |
| 10 | 3<br>4<br>5<br>6 | 480000<br>480000<br>480000<br>480000 | 86<br>86<br>86<br>86 | 5581:1<br>5581:1<br>5581:1<br>5581:1 | 0.89<br>0.82<br>0.83<br>0.83 | 15.80<br>14.60<br>13.55<br>13.93 | 0.66<br>0.51<br>0.65<br>0.51 |

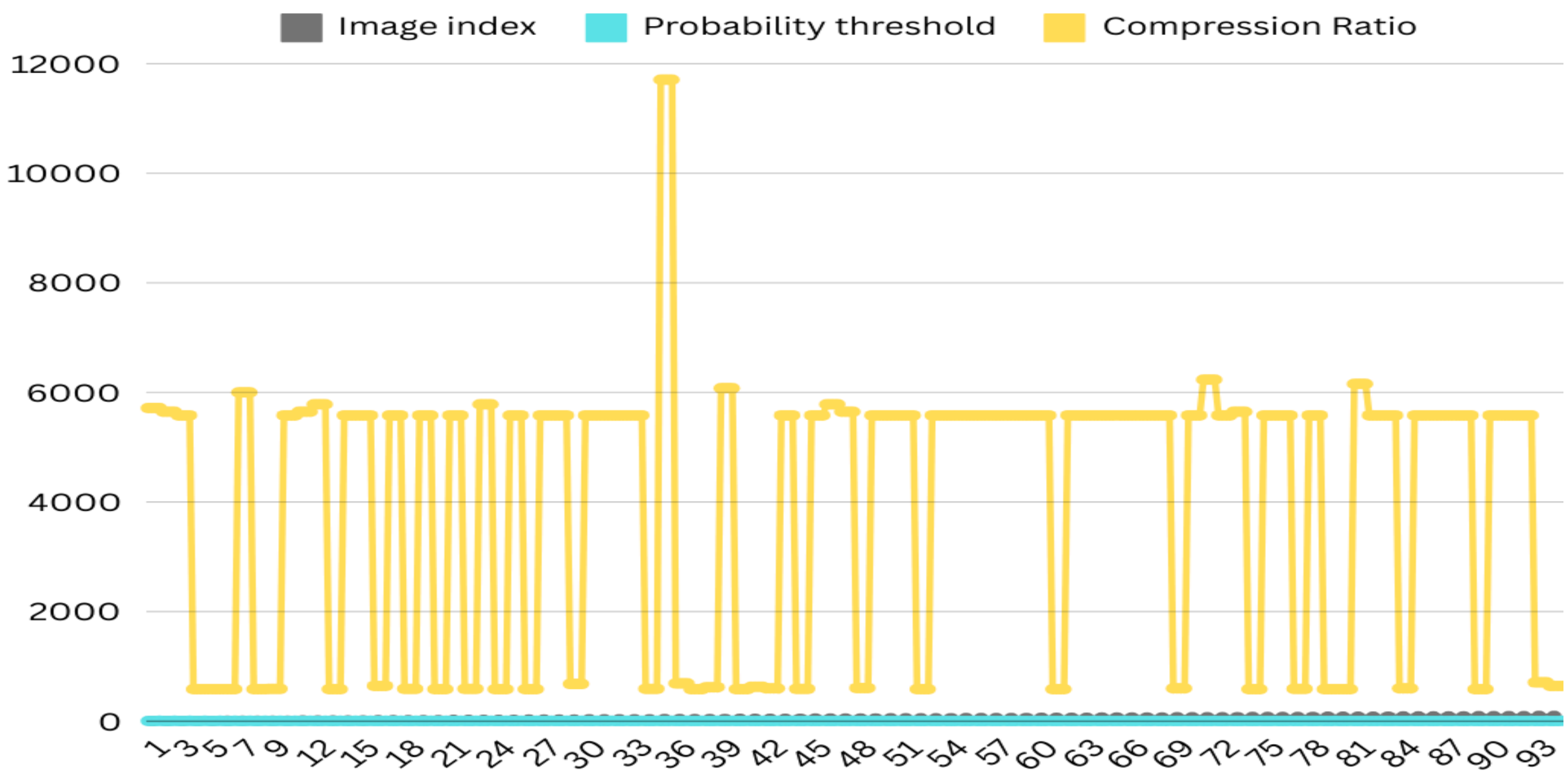

**Fig. 11: The line graph displaying the compression ratio against the probability thresholds 3, 4, 5, 6 for the first 94 images of the dataset.**

**Table 9: Arithmetic Coding Using Cardinality Reduction: Metrics for 10 Dataset Images**

| Image Index | Compression Ratio | RMSE | CPU Usage(%) | Time (s) | Original Size(bytes) | Compressed Size(bytes) |
|---|---|---|---|---|---|---|
| image001.jpg | 3120:1 | 103.75 | 17.8 | 0.0381 | 65536 | 21 |
| image002.jpg | 2979:1 | 103.05 | 7.0 | 0.0161 | 65536 | 22 |
| image003.jpg | 3120:1 | 123.31 | 22.7 | 0.0181 | 65536 | 21 |
| image007.jpg | 2979:1 | 135.04 | 9.4 | 0.0221 | 65536 | 22 |
| image010.jpg | 21845:1 | 165.25 | 16.4 | 0.0182 | 65536 | 3 |
| image011.jpg | 3120:1 | 171.99 | 29.2 | 0.0163 | 65536 | 21 |
| image013.jpg | 3120:1 | 129.74 | 22.3 | 0.0248 | 65536 | 21 |
| image015.jpg | 2979:1 | 70.85 | 49.6 | 0.0213 | 65536 | 22 |
| image016.jpg | 3449:1 | 42.90 | 4.7 | 0.0177 | 65536 | 19 |
| image018.jpg | 2849:1 | 146.91 | 35.7 | 0.0356 | 65536 | 23 |

### 5 Conclusion and Future Work

The article explored Arithmetic coding as an effective compression technique for resource-constrained IoT environments and overcoming its drawbacks for adoption in the IoT environment. Conversion of the images to grayscale resulted in a high compression ratio and low latency with minimal loss in data integrity. By applying the concept of iterative optimization and reducing the dimensionality for faster processing and low computational overhead, the proposed modifications demonstrated significant improvements by achieving a high compression ratio without degrading the data quality, minimizing memory and CPU usage, and faster compression time. The experimental analysis revealed that restricting the probability threshold to 3 decimal places, grouping similar symbols, iterative refinement, and dimensionality reduction can help reduce the computational complexity of the Arithmetic Coding. Compared to the existing approaches, the proposed algorithm was better regarding the compression ratio and the time taken to perform the compression. The approach can be extended to real-time datasets compatible with heterogeneous data types as part of future work. Analyzing the compression algorithm for security is yet another dimension to be explored as part of future work that can guarantee both the efficiency and security of the algorithm.

**Table 10: Comparison with related works on Compression Ratio and Time (*na = not available).**

| Metrics/Method | [35]N-gram +AC | [36]Adaptive AC | [53]Bi-LSTM +AC | Our Proposed work |
|---|---|---|---|---|
| Compression Ratio | 5.6:1 | 4.97:1 | 4.06:1 | 814:1 |
| Time (ms) | 325 | *na | *na | 101 |
| Dataset | Text data | UW Greyset(images) | Power grid data | Image data |
| Time Complexity | $O(n)$ | $O(kn)$ | $O(n^2)$ | $O(n)$ |
| Space Complexity | $O(n)$ | $O(kn)$ | $O(kn)$ | $O(n)$ |